\documentclass{aastex}
\usepackage{emulateapj5,apjfonts}

\slugcomment{2000 April 4, Accepted for publication in The Astrophysical Journal Letters}

\shorttitle{NARROW X-RAY ABSORPTION LINES FROM NGC\,3783}

\shortauthors{KASPI ET AL.}

\begin{document}

\title{Discovery of Narrow X-ray Absorption Lines from NGC\,3783 \\
with the {\it Chandra} HETGS}
\author{
Shai Kaspi,\altaffilmark{1} 
W. N. Brandt,\altaffilmark{1} 
Hagai Netzer,\altaffilmark{2} 
Rita Sambruna,\altaffilmark{1}
George Chartas,\altaffilmark{1} \\
Gordon P. Garmire,\altaffilmark{1} 
and
John A. Nousek\altaffilmark{1}
}
\altaffiltext{1}{Department of Astronomy and Astrophysics, 525 Davey
Laboratory, The Pennsylvania State University, University Park, PA, 16802.}
\altaffiltext{2}{School of Physics and Astronomy and the Wise
Observatory, Raymond and Beverly Sackler Faculty of Exact Sciences, \\
\hglue 0.5cm Tel-Aviv University, Tel-Aviv 69978, Israel.}

\begin{abstract}

We present the first grating-resolution X-ray spectra of the Seyfert~1
galaxy NGC\,3783, obtained with the High Energy Transmission Grating
Spectrometer on the {\it Chandra} X-ray Observatory. These spectra
reveal many narrow absorption lines from the H-like and He-like ions of
O, Ne, Mg, Si, S and Ar, as well as \ion{Fe}{17}--\ion{Fe}{21} L-shell
lines. We have also identified several weak emission lines, mainly from
O and Ne. The absorption lines are blueshifted by a mean velocity of
$\approx 440\pm 200$~km\,s$^{-1}$ and are not resolved, indicating a
velocity dispersion within the absorbing gas of a few hundred
km\,s$^{-1}$ or less. We measure the lines' equivalent widths and
compare them with the predictions of photoionization models. The
best-fitting model has a microturbulence velocity of 150~km\,s$^{-1}$
and a hydrogen column density of $1.3\times 10^{22}$~cm$^{-2}$. The
measured blueshifts and inferred velocity dispersions of the X-ray
absorption lines are consistent with those of the strongest UV
absorption lines observed in this object. However, simple models that
propose to strictly unify the X-ray and UV absorbers have difficulty
explaining simultaneously the X-ray and UV absorption line strengths.

\end{abstract}


\section{Introduction}

X-ray spectra of Active Galactic Nuclei (AGNs) often show evidence for
deep \ion{O}{7} (739~eV) and \ion{O}{8} (871~eV) absorption edges. The
ionized gas component creating these edges is named the `warm absorber'
and is seen in most ($\gtrsim$70\%) Seyfert~1s and some quasars.
Theoretical models for the warm absorber suggest it should also be a
source of X-ray emission (e.g., Netzer 1996) and absorption (e.g.,
Nicastro, Fiore, \& Matt 1999) lines. While the current X-ray data are
consistent with the presence of such lines (e.g., George et~al 1998a),
it has been impossible to study them in any detail.

The bright Seyfert~1 galaxy NGC\,3783 ($V\approx13.5$ mag, $z=0.0097$)
has one of the strongest X-ray warm absorbers known with a column
density of ionized gas of $N_{\rm H}\approx$~(1--2)$\times
10^{22}$~cm$^{-2}$. {\it ASCA} spectra also show warm absorber
variability and excess flux around 600~eV which has been interpreted as
emission lines from the warm absorber, particularly \ion{O}{7}~568~eV
(e.g., George et~al. 1998a, 1998b). UV spectra of NGC\,3783 show
intrinsic absorption features due to \ion{C}{4}, \ion{N}{5} and
\ion{H}{1}. Two kinematic absorption components are known with radial
velocities of $\approx -560$~km\,s$^{-1}$ and $\approx
-1420$~km\,s$^{-1}$ relative to the optical redshift from [\ion{O}{3}].
The \ion{C}{4} absorption shows extreme variability (e.g., Crenshaw
et~al. 1999 and references therein). It is not clear whether the UV and
X-ray absorbing gas are the same, although some relation seems likely
(e.g., Shields \& Hamann 1997).

This letter describes the discovery of a large number of X-ray
absorption lines in the {\it Chandra} spectrum of NGC 3783. In \S~2 we
describe the observation and data analysis, and \S~3 gives the
preliminary interpretation of the results.
\vspace{1cm}

\section{Observation and Data Analysis}
\label{obser}

NGC\,3783 was observed using the High Energy Transmission Grating
Spectrometer (HETGS; C. R. Canizares et al., in preparation) on the
{\it Chandra} X-ray Observatory\footnote{See {\it The Chandra
Proposers' Observatory Guide} at http://asc.harvard.edu/udocs/docs/} on
2000 January 21. The detector was the Advanced CCD Imaging Spectrometer
(ACIS; G. P. Garmire et al., in preparation). The total integration
time was 56~ks, and the observation was continuous. The data were
reduced with the standard pipeline by the {\it Chandra} X-ray Center
(CXC) on 2000 January 26. We used the {\it Chandra} Interactive
Analysis of Observations (CIAO) software to repeat the reduction
process using an updated response matrix file. Our reduction produced
somewhat less noisy spectra than those produced by the pipeline.

The HETGS produces a zeroth order X-ray spectrum at the aim point on
the CCD and higher order spectra which have much higher spectral
resolution along the ACIS-S array. The higher order spectra are from
two grating assemblies, the High Energy Grating (HEG) and Medium Energy
Grating (MEG). Each grating assembly produces two spectra for each
order. Order overlaps are discriminated by the intrinsic energy
resolution of ACIS. The zeroth-order spectrum of NGC~3783 shows
substantial photon pile up, and we leave the detailed analysis of these
data to a future paper. We briefly comment, however, on the presence of
a weak ($\approx 10$ photon) off-nuclear ($\approx 34\arcsec$) source
located at $\alpha_{2000}=$~11$^{\rm h}$39$^{\rm m}$01$\fs$1,
$\delta_{2000}=$~$-37\arcdeg$44$\arcmin$52$\farcs$9, coincident with
one of the galaxy's spiral arms. This source may be a luminous
($L_{0.5-8~{\rm keV}}$ $\approx 2 \times 10^{39}$ ergs\,s$^{-1}$) X-ray
binary in NGC~3783, although it could also be an unrelated background
source.

\begin{figure*}
\centerline{\epsfxsize=18.5cm\epsfbox{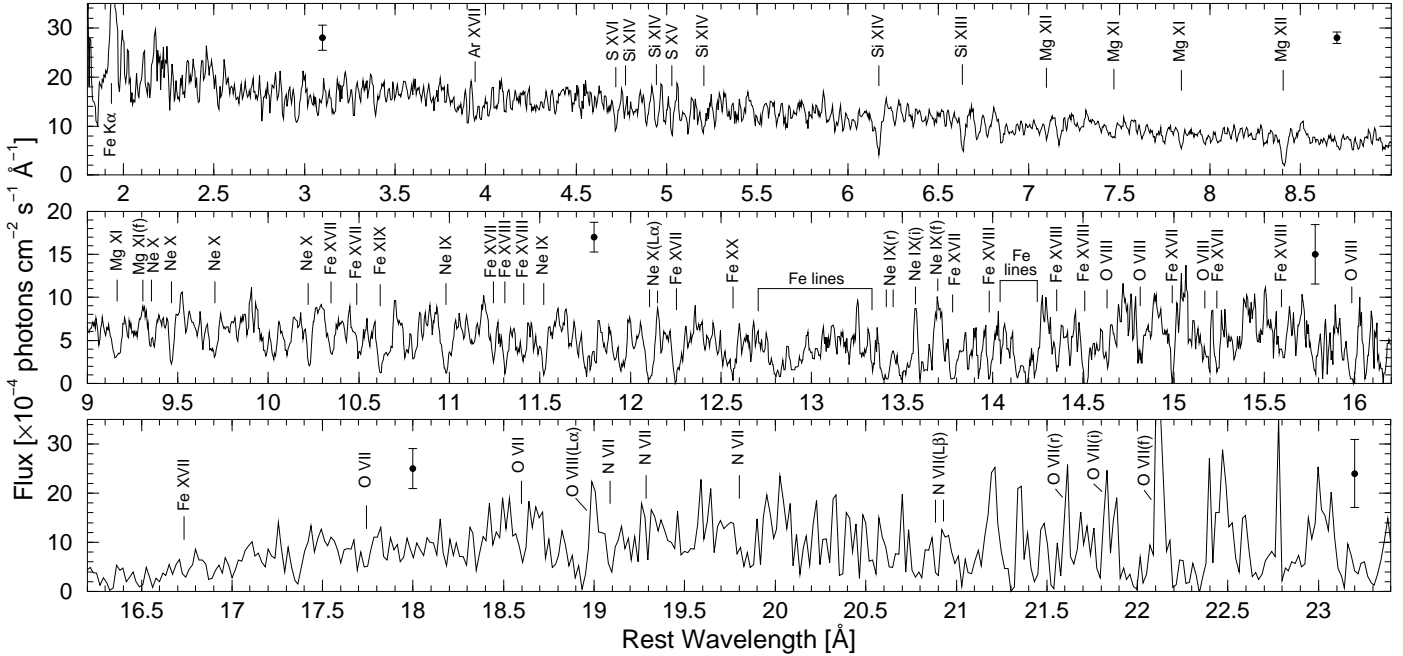}}
\caption{MEG first-order smoothed spectrum of NGC\,3783. The two upper
panels are binned by 0.005~\AA , and the bottom panel is binned by
0.02~\AA . Fluxes are corrected for Galactic absorption. Dots with
error bars show the typical 2$\sigma$ statistical error at various
wavelengths. The identified absorption features are marked. The few
identified emission lines are marked with (r) for resonance, (i) for
intercombination, and (f) for forbidden for the He-like ions, as well as
L$\alpha$ and L$\beta$ for the H-like ions.}
\label{megspec}
\end{figure*}

The first-order spectra from the HEG and MEG have S/N ratios of
$\approx 5$ and $\approx 2.5$, respectively (at around 7~\AA ). The
higher order spectra had only a few ($\approx 3$) photons per bin and
will not be presented here. In order to flux calibrate the spectra we
used the CIAO software to produce Ancillary Response Files (ARFs). The
flux calibration error is estimated to be $\lesssim$10\% below 6~\AA\
and $\lesssim$20\% above 6~\AA\ (CXC, private communication). We also
corrected the spectra for the Galactic absorption of $N_{\rm
H}=8.7\times 10^{20}$~cm$^{-2}$ and the cosmological redshift.

The two MEG first-order spectra agree well with each other. The
dispersions of the MEG spectra are 0.005~\AA\ per bin, and the spectral
resolution is $\approx 0.023$~\AA\ (approximately constant across the
entire HETGS band). The two HEG spectra also agree well with each other
and with the MEG spectra, though they have lower S/N. The HEG spectra
have dispersions of 0.0025~\AA\ per bin, and the spectral resolution is
about two times better than for the MEG spectra. In Fig.~\ref{megspec}
we present the mean of the two MEG spectra. This mean spectrum was
smoothed using a simple 3-bin boxcar filter. Since above
$\sim$16~\AA\ there are $\lesssim$4 counts per 0.005~\AA\ bin, the
spectrum presented in the bottom panel of Fig.~\ref{megspec} is binned
to 0.02~\AA .

The overall shape and flux of the X-ray spectrum are consistent with
those found in previous X-ray observations of NGC\,3783 (e.g., George
et~al. 1998b). The {\it Chandra} fluxes in several energy bands are
also consistent with those from a simultaneous 4~ks {\it ASCA} spectrum
to within $\approx$~10\%. The total flux measured from the {\it
Chandra} spectrum is
(2.0$\pm$0.2)$\times$10$^{-11}$~ergs\,cm$^{-2}$\,s$^{-1}$ in the
0.5--2~keV band and
(4.2$\pm$0.9)$\times$10$^{-11}$~ergs\,cm$^{-2}$\,s$^{-1}$ in the
2--7~keV band. When plotting the modeled {\it ASCA} spectrum on top of
the {\it Chandra} spectrum, we find general consistency in the overall
shape as well as with the oxygen absorption edges.

The most striking new features seen in the {\it Chandra} spectrum of
NGC\,3783 are narrow X-ray absorption lines. In order to measure these
spectral features we interpolated over all the absorption lines and
fitted the resulting continuum with a cubic spline which serves as the
continuum reference point for measuring the lines. In Table~1
we list the information about selected absorption and emission lines.
In Fig.~\ref{megspec} we identify some absorption lines which are not
listed in Table~1 due to brevity and lack of model
predictions. There are also some absorption and emission features which
appear to be present but are not identified yet. We used multiple
lines from the same ion to help with line identification. We checked
for the existence of these lines in each of the MEG spectra (and also
in the HEG spectra when possible) and measured them in the smoothed
average spectrum.

The measured lines' Full Widths at Half Maximum (FWHMs) are consistent
with the MEG resolution which is $\approx 1300$~km\,s$^{-1}$ at
5~\AA\ and $\approx 500$~km\,s$^{-1}$ at 15~\AA . Hence, the lines are
not clearly resolved (see discussion in \S~\ref{dinamic}). We were not
able to put better constraints on the velocity field with the HEG due
to the small S/N.

\section{Discussion}

\subsection{Absorption Lines, Column Density and Ionization Parameter}
\label{ionsec}

The strongest observed absorption lines in the X-ray spectrum of
NGC\,3783 are from the H-like and He-like ions of O, Ne, Mg, Si, S and
Ar, as well as \ion{Fe}{17}--\ion{Fe}{21} L-shell lines (see
Table~1 but note that some features represent blends, and the
values given are the summed EWs).

As explained above, the lines are not resolved, and the only firm
conclusion is that, given a Gaussian profile, their FWHMs do not exceed
a few hundred km\,s$^{-1}$. Given this uncertainty, we have taken an
approach that relies solely on the measured EWs of the lines. We
assumed a certain microturbulence velocity in a single line-of-sight
cloud and calculated a theoretical spectrum with the photoionization
code ION99, \ the 1999 \ version of the  

%\begin{deluxetable}{lcccc}
%\tablecolumns{5}
%\tabletypesize{\small}
%\tabletypesize{\footnotesize}
%\tabletypesize{\scriptsize}
%\tablewidth{0pt}
%\tablecaption{Absorption and Emission Lines From NGC\,3783
%\label{lines}}
%\tablehead{
%
\footnotesize
\begin{center}
{\sc TABLE 1\\
Absorption and Emission Lines From NGC\,3783}
\vskip 4pt
\begin{tabular}{lcccc}
\hline
\hline
{} &
{Predicted} &
{Velocity} &
{Predicted} &
{Measured} \\
{} &
{$\lambda$} &
{Shift} &
{EW} &
{EW} \\
{Ion} &
{(\AA)} &
{(km\,s$^{-1}$)} &
{(m\AA)} &
{(m\AA)} \\
\hline
\multicolumn{5}{c}{Absorption Lines} \\
\hline
\ion{S }{16}                  &  4.728 & $-$571$\pm$317          &   0.7 &   8.0$\pm$2.6 \\ 
\ion{S }{15}                  &  5.039 & $-$535$\pm$297          &   8.2 &   7.5$\pm$2.5 \\ 
\ion{Si}{14}                  &  5.217 & $-$920$\pm$287          &   1.4 &  10.1$\pm$3.5\phn \\ 
\ion{Si}{14}                  &  6.181 & $-$534$\pm$242          &   7.7 &  17.3$\pm$4.6\phn \\ 
\ion{Si}{13}                  &  6.648 & $-$586$\pm$225          &  18.4 &  14.7$\pm$4.8\phn \\ 
\ion{Mg}{12}                  &  7.106 & $-$379$\pm$211          &   6.6 &   6.2$\pm$2.2 \\ 
\ion{Mg}{11}                  &  7.473 & $-$280$\pm$200          &   8.6 &   8.0$\pm$3.6 \\ 
\ion{Mg}{11}                  &  7.851 & $-$382$\pm$191          &  15.6 &   9.7$\pm$3.1 \\ 
\ion{Mg}{12}\tablenotemark{a} &  8.420 & $-$463$\pm$178          &  $>$19.9 &  25.6$\pm$6.6\phn \\ 
\ion{Mg}{11}\tablenotemark{a} &  9.170 & $-$261$\pm$163          &  $>$25.9 &  34.5$\pm$10.2 \\ 
\ion{Ne}{10}                  &  9.362 & $-$384$\pm$320          &   6.2 &  16.4$\pm$4.5\phn \\ 
\ion{Ne}{10}\tablenotemark{b} &  9.481 & $-$474$\pm$158          &  12.0 &  18.6$\pm$4.9\phn \\ 
\ion{Ne}{10}\tablenotemark{a} &  9.709 & $-$216$\pm$154          &  $>$15.7 &  29.7$\pm$8.2\phn \\ 
\ion{Ne}{10}\tablenotemark{a} & 10.239 & $-$410$\pm$146          &  $>$24.7 &  19.8$\pm$5.5\phn \\ 
\ion{Ne}{ 9}\tablenotemark{a} & 11.001 & $-$518$\pm$190          &  $>$20.1 &  50.4$\pm$12.2 \\ 
\ion{Fe}{17}                  & 11.250 & $-$186$\pm$133          &  23.3 &  15.7$\pm$4.3\phn \\ 
\ion{Fe}{18}\tablenotemark{a} & 11.435 & $-$629$\pm$131          &  $>$20.5 &  21.0$\pm$14.2 \\ 
\ion{Ne}{ 9}\tablenotemark{a} & 11.547 & $-$701$\pm$129          &  $>$49.3 &  27.2$\pm$7.4\phn \\ 
\ion{Ne}{10}\tablenotemark{b} & 12.132 & $-$667$\pm$123          &  69.8 &  49.2$\pm$19.8 \\ 
\ion{Fe}{17}\tablenotemark{b} & 12.263 & $-$391$\pm$122          &  43.1 &  51.9$\pm$12.8 \\ 
\ion{O }{ 8}                  & 14.635 & \phn$-$20$\pm$102       &  34.3 &  24.8$\pm$12.2 \\ 
\ion{O }{ 8}                  & 14.821 & $-$182$\pm$101          &  38.9 &  27.8$\pm$12.8 \\ 
\ion{O }{ 8}                  & 15.176 & \phn$-$79$\pm$197       &  42.8 &  35.8$\pm$11.2 \\ 
\ion{Fe}{17}                  & 15.264 & $-$471$\pm$196          &  42.7 &  30.6$\pm$8.4\phn \\ 
\ion{Fe}{18}                  & 15.628 & $-$624$\pm$249          &  18.2 &  27.7$\pm$7.9\phn \\ 
\ion{O }{ 8}\tablenotemark{a} & 16.007 & $-$346$\pm$121          &  $>$52.3 &  43.9$\pm$14.2 \\ 
\ion{O }{ 7}                  & 18.628 & $-$515$\pm$322          &  52.1 &  41.9$\pm$14.8 \\ 
\ion{N }{ 7}                  & 19.120 & $-$510$\pm$109          &  11.2 &  18.8$\pm$6.8\phn \\
\hline
\multicolumn{5}{c}{Emission Lines} \\
\hline
\ion{Mg}{12}                  &  9.300 & \phm{$-$}306$\pm$225    &  15 &  11.0$\pm$6.8\phn \\ 
\ion{Ne}{10}                  & 12.132 & \phm{$-$}444$\pm$123    &  36 &  44$\pm$24 \\ 
\ion{Ne}{ 9}                  & 13.447 & \phm{$-$}\phn64$\pm$111 &  19 &  99$\pm$88 \\ 
\ion{Ne}{ 9}                  & 13.553 & \phm{$-$}464$\pm$110    &  13 & 163$\pm$140 \\ 
\ion{Ne}{ 9}                  & 13.699 & \phm{$-1$}26$\pm$175     &  41 & 245$\pm$215 \\ 
\ion{O }{ 8}                  & 18.967 & \phm{$-$}415$\pm$126    & 460 & 335$\pm$310 \\ 
\ion{N }{ 7}                  & 20.912 & \phm{$-$}161$\pm$100    &  28 & 118$\pm$92\phn \\ 
\ion{O }{ 7}                  & 21.602 & \phm{$-1$}54$\pm$69\phn &  99 & 197$\pm$169 \\ 
\ion{O }{ 7}                  & 21.807 & \phm{$-$}229$\pm$68\phn &  52 & 135$\pm$109 \\ 
\ion{O }{ 7}                  & 22.102 & \phm{$-$}154$\pm$67\phn & 201 & 631$\pm$583 \\ 
\hline
\end{tabular}
\vskip 2pt
\parbox{3.485in}{
\small\baselineskip 9pt
\footnotesize
\indent
$\rm ^a${Blended\,with\,Fe\,absorption lines which\,are\,not\,taken into
account in the model.} \\
$\rm ^b${Blended with Fe absorption lines which are taken into account
in the model.}
}
\end{center}
\setcounter{table}{1}
\normalsize

\noindent code ION (Netzer 1996). We then
compared the results with the observations. The most important
ingredients of the modelare the microturbulence velocity, the column
density, the spectral energy distribution (SED), the ionization
parameter, and the gas composition.

The atomic data required for calculating line optical depths are taken
from various sources. The oscillator strengths for the H-like and
He-like transitions of all metals are well known (e.g., Porquet \&
Dubau 2000 and references therein). This is not the case for the
various iron L-shell lines where there is no standard published data
set. In this paper we rely on various publications (e.g., Mason et al.
1979; Bhatia \& Mason 1980; Cornille et al. 1992, 1994; Phillips et
al. 1996; Saba et al. 1999) as well as on the National Institute of
Standards and Technology (NIST) data set. These were combined with the
Phillips et~al. (1999) line lists, which are based on solar X-ray
spectra and served as our major source for the wavelengths of the iron
lines. Only 3--5 of the strongest iron lines, per ion, have been used
in the analysis.

Our models are based on those in George et~al. (1998b) that gave
satisfactory fits to the 1993 and 1996 {\it ASCA} observations of the
source. They include a weak UV-bump AGN SED (the weak IR case in Netzer
1996) with $\Gamma_{0.1-50~{\rm keV}}$=1.75, a single constant density
cloud with $N_{\rm H}=1.3\times 10^{22}$~cm$^{-2}$, solar com-
%
%\begin{figure}
\centerline{\epsfxsize=8.5cm\epsfbox{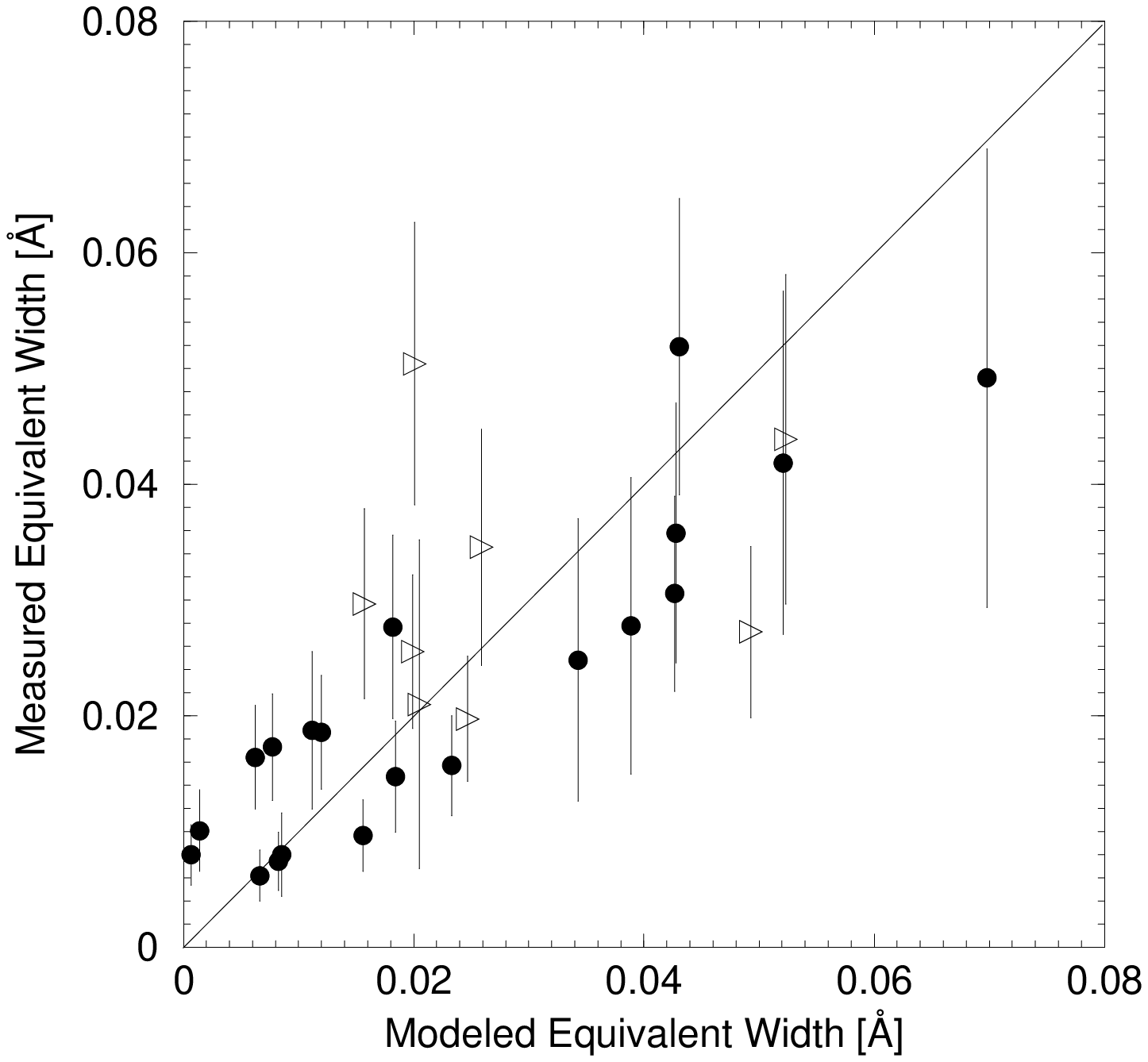}}
\figcaption{Measured versus modeled absorption line EWs. Filled circles
are model EWs that are a sum of all lines contributing to the blend.
Right-pointing triangles are lower limits for model EWs (see
\S~\ref{obser} for details). The agreement between the model and the
measured points is noticeable (Spearman rank-order correlation
probability of $\approx 10^{-6}$). A line with a slope of unity is
drawn to guide the eye.
\label{ewcomp}}
%\end{figure}
\centerline{}

%\begin{deluxetable}{lclc}
%\tablecolumns{4}
%\tabletypesize{\small}
%\tabletypesize{\footnotesize}
%\tabletypesize{\scriptsize}
%\tablewidth{0pt}
%\tablecaption{Best Model Column Densities
%\label{colden} }
%\tablehead{

\footnotesize
\begin{center}
{\sc TABLE 2\\
Best Model Column Densities}
\vskip 4pt
\begin{tabular}{lclc}
\hline
\hline
{} &
{Column} &
{} &
{Column} \\
{Ion} &
{($10^{17}$ cm$^{-2}$)} &
{Ion} &
{($10^{17}$ cm$^{-2}$)} \\
\hline
\ion{O}{7} &    5.0 & \ion{S}{16} &  0.05   \\
\ion{O}{8}&     38 & \ion{Ar}{17} &  0.06    \\
\ion{Ne}{9}&  4.0  & \ion{Ar}{18} & 0.003    \\
\ion{Ne}{10}& 6.9  & \ion{Fe}{17} & 1.2     \\
\ion{Mg}{11}&  2.5  & \ion{Fe}{18} & 1.5    \\
\ion{Mg}{12} &  1.5 & \ion{Fe}{19} & 1.0   \\
\ion{Si}{13} &  2.2 & \ion{Fe}{20} & 0.4   \\
\ion{Si}{14} &  0.5 & \ion{Fe}{21} & 0.07  \\
\ion{S}{15} & 0.5  & \ion{Fe}{22} & 0.02   \\
\ion{C}{4} & 2.4$\times 10^{-5}$ &  \ion{N}{5} & 4.7$\times 10^{-4}$  \\ 
\hline
\end{tabular}
\end{center}
\setcounter{table}{2}
\normalsize
\centerline{}

\noindent position,
and a 0.1--10 keV ionization parameter of $U_{\rm x}=0.13$. The
standard ($E>13.6$~eV) ionization parameter for this continuum is 19.3
and the mean temperature is 1.9$\times 10^5$~K. We have assumed
Doppler profiles with $V^2=V^2_{\rm thermal}+V^2_{\rm turbulence}$ and
calculated absorption EWs for $100<V_{\rm
turbulence}<600$~km\,s$^{-1}$. The best agreement with observations is
for $V_{\rm turbulence}$=150~km\,s$^{-1}$. The model EWs are tabulated
in Table~1. For a few lines we list only the model's EW
lower-limit since the measurement includes blends for which we do not
yet have a modeled EW. In most cases, the blends are expected to
increase the modeled EW by $\approx$~10--20\%. The comparison between
the observed and calculated EWs is shown in Fig.~\ref{ewcomp}, and the
deduced column densities, for the ions producing the strongest lines,
are given in Table~2. All of the strong lines predicted by
the model are observed in the spectra. The microturbulence velocity of
our best fitted model is consistent with the velocity dispersion of the
UV absorption lines (161$\pm$22~km\,s$^{-1}$; Crenshaw et~al. 1999).

We have carried out basic tests to verify that the deduced line widths
are in the right range. If the `true' widths are much larger than
assumed, the resonance lines in the H-like and He-like series will be
optically thin. The observed EWs show that this is not the case. On the
other hand, if the `true' line widths are much smaller than
150~km\,s$^{-1}$, several of these lines will be saturated, giving very
similar EWs to all. This is not observed either. While these are not
strong conclusions given the S/N of the observations, it is encouraging
to see the good agreement between the observed and calculated EWs,
suggesting that the line optical depths are indeed in the intermediate
range. According to the model, the largest optical depth for a line
with $E>1$~keV is $\tau$(\ion{Ne}{10}~12.132~\AA)=35.

\subsection{Dynamics of the Absorbing Gas}
\label{dinamic}

All the identified absorption lines in the HETG spectrum of NGC\,3783
are blueshifted by a few hundred~km\,s$^{-1}$ with respect to the
systemic velocity. The unweighted mean blueshift is
440$\pm$200~km\,s$^{-1}$. Kaastra et~al. (2000) have recently observed
the bright Seyfert~1 NGC\,5548 using the Low Energy Transmission
Grating Spectrometer (LETGS) on {\it Chandra}. The low-energy X-ray
lines reported in that paper indicate a similar level of ionization,
and the absorption lines are blueshifted by similar amounts. Judging by
these first two results, it seems that the warm X-ray material in
Seyfert~1 galaxies is outflowing with typical velocities of several
hundred km\,s$^{-1}$. This has important implications for the dynamics
of the absorbing gas as well as for the X-ray-UV connection.

The acceleration of X-ray ionized gas by the strong central radiation
source in AGN has not been studied in detail so far (for some
discussion see Reynolds \& Fabian 1995). The recent work by Chelouche
\& Netzer (2000) is the first detailed study of this kind. This study
discusses the motion of discrete, X-ray ionized clouds in hydrostatic
equilibrium, under a large range of conditions. It shows that
acceleration to velocities of a few hundred to a few thousand
km\,s$^{-1}$ is a natural consequence of the AGN environment. The
terminal velocity is of the same order as the escape velocity in the
region where the flow originates. Using these results, and assuming
that the conditions in the nucleus of NGC\,3783 are similar to those
considered by Chelouche \& Netzer (2000), we conclude that the origin
of the warm absorbing gas in this source is outside the Broad Line
Region, where the escape velocity drops below $\approx
1000$~km\,s$^{-1}$. Better line profile analysis will enable us to
constrain this location more accurately.

Another important comparison is with the UV absorption lines observed
from this source. Several papers (e.g., Mathur, Elvis, \& Wilkes 1995;
Shields \& Hamann 1997 and references therein) suggest that the UV and
X-ray absorbers may originate in the same component. The Crenshaw et
al. (1999) observations reveal two UV absorption systems, the strongest
of which has a velocity consistent with the mean observed velocity of
the X-ray gas and a velocity dispersion consistent with our best fitted
model. Our model, which depends on the assumed SED, predicts column
densities for \ion{C}{4} and \ion{N}{5} (see Table~2) which
are 37 and 5 times smaller, respectively, than what has been calculated
by Crenshaw et al. (1999) using the 1993--1995 data. Hence the UV
absorption lines appear to be inconsistent with our X-ray absorber
model and are likely coming from a separate component.

\subsection{Emission Lines and Covering Factor}

The HETG spectra of NGC\,3783 revel several weak emission lines (see
Table~1) some of which show hints of P Cygni profiles. These
have a marginally significant redshift with an unweighted mean of
230$\pm$170~km\,s$^{-1}$ relative to the systemic velocity measured
from the [\ion{O}{3}] line. Kaastra et~al. (2000) found similar
redshifts for the emission lines in the LETG spectrum of NGC\,5548.
While the EWs of some of the lines are well within the MEG and HEG
capability, the low S/N and bumpy continuum do not allow reliable EW
measurements. The values given in Table~1 are therefore
highly uncertain. However, even these rough measurements are good
enough to test the idea that the central source in NGC\,3783 is
surrounded by gas clouds with properties similar to those deduced for
the line-of-sight absorber. This is done by comparing the EWs
calculated by ION with those observed, for various assumed covering
factors.

The calculated emission-line EWs, for an assumed geometry of full
covering, are listed in Table~1. Given these values, we find
that the mean deduced covering factor is close to unity, in agreement
with previous estimates (e.g., George et al. 1998b). This is the {\it
first direct constraint} on the covering factor of a warm absorber in
any AGN. Unfortunately, the uncertainty on the covering factor is
large, due to the uncertain EWs. Further discussion of this issue, as
well as detailed analysis of the continuum, emission features, and
absorption features, are deferred to a future paper.

\acknowledgments
We thank all the members of the {\it Chandra} team for their enormous
efforts. We also thank Herman L. Marshall and an anonymous referee for
helpful comments. We gratefully acknowledge the financial support of
NASA grant NAS~8-38252 (G. P. G., PI), NASA LTSA grant NAG~5-8107 (S.
K., W. N. B.), and the Alfred P. Sloan Foundation (W. N. B.). H. N.
acknowledges support from the Israel Science Foundation and the Jack
Adler Chair for Extragalactic Astronomy.

\end{document}